# WATERMARKING AND ANOMALY DETECTION IN MACHINE LEARNING MODELS FOR LORA RF FINGERPRINTING

*Aarushi Mahajan*[1], *Wayne Burleson*[2]

[1]University of Massachusetts Amherst, aarushimahaj@umass.edu
[2]University of Massachusetts Amherst, burleson@umass.edu

## ABSTRACT

Radio frequency fingerprint identification (RFFI) distinguishes wireless devices by the small variations in their analog circuits, avoiding heavy cryptographic authentication. While deep learning on spectrograms improves accuracy, models remain vulnerable to copying, tampering, and evasion. We present a stronger RFFI system combining watermarking for ownership proof and anomaly detection for spotting suspicious inputs. Using a ResNet-34 on log-Mel spectrograms, we embed three watermarks: a simple trigger, an adversarially trained trigger robust to noise and filtering, and a hidden gradient/weight signature. A convolutional Variational Autoencoders (VAE) with Kullback-Leibler (KL) warm-up and free-bits flags off-distribution queries. On the LoRa dataset, our system achieves 94.6% accuracy, 98% watermark success, and 0.94 AUROC, offering verifiable, tamper-resistant authentication.

*Index Terms*— Anomaly detection, Autoencoders, Deep learning, Model watermarking, Radio frequency fingerprint identification (RFFI), ResNet, Spectrograms, Wireless device authentication

## 1. INTRODUCTION

Radio frequency fingerprint identification (RFFI) is becoming an attractive option for authenticating IoT devices because it avoids the overhead of cryptographic methods[1, 3]. Using deep learning on spectrograms increases resilience to channel fluctuations [2, 5], yet most prior work remains focused on accuracy and ignores adversarial threats. In practice, stolen models can be redistributed, pruned, or fine-tuned (*weight tampering*), while device fingerprint can be suppressed by adversarial perturbations or input evasion [4, 7].

We propose a defense-in-depth approach to RFFI. The input signals are first converted into spectrograms with a fixed STFT–Mel front end and then classified using a ResNet-34 model, which offers both high accuracy and support for watermarking. We add three types of watermarks that protect the model against copying, input evasion, and weight modifications. In addition, a convolutional VAE acts as a guard that flags inputs which do not match normal device patterns. On the LoRa dataset, our system reaches 94.6% classification accuracy, over 98% success in watermark verification, and an AUROC of 0.94 for anomaly detection. These results show that RFFI can be made not only accurate but also reliable and secure, even when facing adversarial conditions.

## 2. RELATED WORK

Early work showed that CNNs based on spectrograms could reliably identify LoRa devices [1], and later research added augmentation and normalization to make them more robust to channel effects such as carrier frequency offset (CFO) and fading [5]. These approaches set strong baselines, however, they failed to account for risks such as adversarial attacks or model theft.

Other research modeled device-level imperfections—like oscillator drift, I/Q imbalance, and power amplifier nonlinearities—to explain why fingerprints are stable across time and hardware [3]. Adversarial training efforts, such as RFAL, highlighted how easy it can be to fool RFFI systems with adversarial inputs [4]. Surveys of deep learning RFFI [2] and multi-domain approaches that combine spectral, temporal, and constellation features [7] pushed accuracy further but still left security questions open. Autoencoders have also been used for spectrum anomaly detection and interference mitigation [8, 9], and newer VAE variants like ED-VAE [6] improve training stability. However, these methods have not been applied to RFFI.

**Threat Model.** In summary, previous work shows that CNNs achieve strong accuracy, adversarial training is needed to handle malicious input, and auto-encoders can detect anomalies. Yet, no prior system consolidates these defenses. We fill this gap by combining layered watermarking with an anomaly guard, enabling secure and verifiable RFFI in adversarial settings.

## 3. METHODOLOGY

### 3.1. Problem Setting and Threat Model

In addition to classifying devices, radio frequency fingerprint identification (RFFI) must protect the model itself. We assume that adversaries can do three things: (i) steal a trained model and use it for their benefit (*model theft*); (ii) modify the model specifications to hide its origin by pruning, quantizing, or fine-tuning (*weight tampering*); and (iii) attempt to evade ownership checks by adding adversarial noise (*input evasion*).

The removal of watermarks through pruning has been studied [10], and similar problems have been observed in adversarial learning for RF identification [4]. Our main goal is to develop an RFFI system that maintains accuracy when there are any channel fluctuations, also incorporating anomaly detection and watermarking to confirm ownership and expose tampering.

### 3.2. Data and Signal Front End

We use LoRa transmissions recorded with a Universal Software Radio Peripheral (USRP) N210, following the dataset in [1]. Each packet preamble is converted to a *log-Mel spectrogram* using a fixed STFT–Mel–log pipeline:

$$X(m, k) = \sum_{n=0}^{N-1} x[n]\, w[n - mR]\, e^{-j2\pi kn/N}, \quad (1)$$

$$E_m = \sum_{k=0}^{N-1} |X(m, k)|^2 H_m(k), \quad (2)$$

$$S(m) = \log(E_m + \varepsilon), \quad (3)$$

where $w$ is the window, $R$ the hop size, $H_m$ the Mel filters, and $\varepsilon > 0$ ensures stability. Spectrograms highlight device-level impairments and suppress noise. Previous surveys [2], [3], and channel-robust LoRa studies [5] confirm their suitability against raw IQ characteristics. This preprocessing stage is the same across all experiments.

### 3.3. Classifier Backbone

**ResNet-34** is our main classifier. Its residual blocks

$$y = F(x; W) + x \quad (4)$$

allow for training deeper networks without sacrificing important device-specific features. Previous research has shown that CNN and ResNet models perform better than shallow networks in narrowband systems [3], scalable LoRa configurations [5], and multi-domain fingerprinting tasks [7]. The architecture of ResNet-34 provides a good balance between accuracy and computational cost, making it perfect for adding watermarking techniques.

In our experiments, ResNet-18 struggled to capture important yet very fine device specific characteristics, particularly when there was channel variability. Its shallow depth limits its expressiveness, resulting in a slightly lower accuracy (91.8% vs. 94.6% for ResNet-34). Hence, ResNet-34 is more suitable for this task.

### 3.4. Training-Integrated Watermarking

We integrate three important watermarking schemes after data preprocessing, that is, spectrogram formation.

**(i) Trigger watermark.** A small subset of spectrograms is modified with a deterministic pattern $T(\cdot)$ and mapped to a reserved class $y_{wm}$:

$$L_{trig} = CE\left(f_\theta(T(S)), y_{wm}\right). \quad (5)$$

This provides a black-box ownership check [7].

**(ii) Adversarial watermark.** To resist data evasion, triggered spectrograms are perturbed within $\|\delta\|_p \leq \epsilon$ and forced to the watermark label:

$$S' = T(S), \quad \tilde{S} = S' + \delta, \quad (6)$$

$$L_{adv} = CE\left(f_\theta(\tilde{S}), y_{wm}\right). \quad (7)$$

This improves robustness against denoising and adversarial attacks, consistent with RFAL [4].

**(iii) Gradient/parameter watermark.** Penultimate features are aligned with a secret vector $v$:

$$L_{sig} = \lambda\left(1 - \frac{\langle \hat{f}, v \rangle}{\|\hat{f}\|\|v\|}\right), \quad \hat{f} = \frac{f_\theta^{pen}(S)}{\|f_\theta^{pen}(S)\|}. \quad (8)$$

This preserves clean accuracy while embedding a signature resilient to pruning and quantization [10].

### 3.5. Model-Agnostic Guard: ConvVAE

We deploy a convolutional VAE trained only on clean spectrograms as an anomaly detector. With encoder $q_\phi(z|S)$ and decoder $p_\theta(S|z)$, the loss is:

$$L_{VAE}(S) = \|S - \hat{S}\|_2^2 + \beta \sum_{d=1} \max(KL_d, \tau_{fb}), \quad (9)$$

where $\beta$ is annealed during training and $\tau_{fb}$ implements free-bits regularization. At inference, inputs are flagged if their reconstruction/ELBO score exceeds a threshold. Unlike prior AE applications in spectrum sensing [9] and interference mitigation [8], our robust ConvVAE—motivated by ELBO decomposition methods [6]—serves as a security guard against distributional shift and sanitized triggers.

The complete system integrates (i) a fixed STFT–Mel front end [1]–[3], [5], (ii) a ResNet-34 classifier [3], [5], [7], (iii) three complementary watermarking schemes [4], [7], [10], and (iv) a robust ConvVAE guard [6], [8], [9]. Collectively, these components deliver reliable RFFI with verifiable ownership under adversarial conditions.

## 4. EXPERIMENTAL SETUP AND RESULTS

### 4.1. Autoencoder Guard: Variants, Training, Scoring

We evaluated auto-encoders (AEs) as anomaly detectors operating on log-Mel spectrograms. All models share the same front-end pipeline: $f_s$=1 MHz, $N_{FFT}$=256, hop size 128, and 32 Mel bands. Training is performed on devices 0–29 with 800 packets per device; validation uses 20% of training; and testing is on the *seen-devices* set (400 packets per device). All AEs are trained exclusively on clean, non-watermarked inputs.

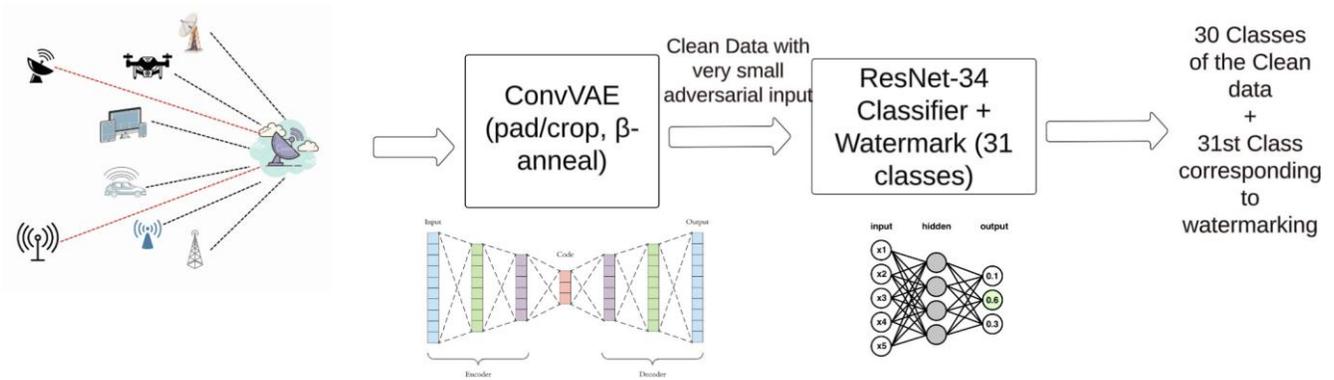

**Fig. 1**. LoRa signals are converted to log-Mel spectrograms and processed by (i) a CAE/ConvVAE guard for anomaly/adversarial detection and (ii) a ResNet-34 classifier with trigger, adversarial, and gradient watermarks. The model outputs 30 device classes plus a reserved watermark class for robust identification and ownership verification.

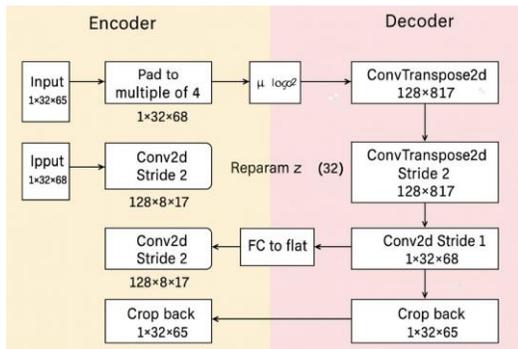

**Fig. 2**. ConvVAE architecture for anomaly detection. Encoder: strided Conv2d layers → latent mean/variance → reparameterization ($z$=32). Decoder: FC projection + ConvTranspose2d layers with crop/pad to restore (1×32×65)

**Variants.**

- **ConvVAE (pad/crop, $\beta$-anneal):** Lightweight CNN VAE (latent size 32, base channels 64), with pad/crop to (32×65) and $\beta$-annealed ELBO training.

- **CAE (U-Net denoising AE):** Symmetric U-Net CAE with skip connections, trained as a denoiser using mild time/frequency masking and additive Gaussian noise with MSE loss.

- **Robust ConvVAE (KL warm-up + free-bits):** Higher-capacity CNN VAE (latent size 64), trained with KL warm-up over the first 10 epochs and free-bits regularization (0.02 nats/dim) to encourage balanced latent utilization.

**Table 1**. Autoencoder Variants for Anomaly Guarding in RFFI (thresholds set for ≈5% clean-data FPR).

| Model | Keep Rate(%) | AUROC | AP |
|---|---|---|---|
| ConvVAE (pad/crop, $\beta$-anneal) | 95.91 | 0.92 | 0.90 |
| CAE (U-Net denoising AE) | 93.80 | 0.89 | 0.87 |
| Robust ConvVAE (KL warm-up + free-bits) | 93.20 | 0.90 | 0.88 |

**Training and Scoring.** Optimization uses AdamW ($3\times10^{-4}$), weight decay $10^{-4}$, batch size 64, and early stopping (patience 5); the robust ConvVAE uses $2\times10^{-3}$ LR and patience 10. At test time, anomaly scores are

$$s(x) = \alpha \, \|x - \hat{x}\|_2^2 + (1 - \alpha)\, \text{ELBO}^-,$$

with thresholds $\tau$ chosen for ≈5% FPR on clean validation data.

### 4.2. Results on Seen-Devices Test

Table 1 shows results. The ConvVAE with pad/crop and $\beta$-annealing achieves the best keep rate (95.9%) and AUROC (0.92). The robust ConvVAE yields balanced AUROC (0.90) and AP (0.88) due to stable latent usage, while the U-Net CAE underperforms (93.8% keep, AUROC 0.89), likely from over-generalization.

**Observations.** The pad/crop ConvVAE offers the best retention and robustness, making it suitable for anomaly guarding under low FPR. U-Net CAEs, while effective for denoising, tend to over-generalize and blur anomalies. KL warm-up with free-bits regularization stabilizes latent capacity and yields balanced AUROC/AP, though with modestly lower keep rate.

**Table 2**. Classifier Backbones on Log–Mel Spectrograms: Accuracy, Trade-offs, and Threat-Model Fit

| Backbone | Acc. (%) | Capacity / Convergence | Channel Robustness | Watermark Compatibility |
|---|---|---|---|---|
| Shallow CNN | 87.4 | Low; may underfit | Limited; heavy augmentation needed | Trigger feasible; parameter marks unstable |
| ResNet-18 | 91.8 | Medium; stable | Better than shallow CNN | Trigger + parameter feasible |
| **ResNet-34 (default)** | **94.6** | High; stable at depth | Good with augmentation | Trigger, adversarial, and parameter marks supported |
| ResNet-35 (custom) | 94.9 | Slightly higher capacity than R-34 | Similar to R-34 | All marks; small extra compute |
| Hybrid feature+distance | 92.1 | Depends on extractor | Scales enrollment | Trigger feasible; parameter marks awkward |

**Table 3**. ResNet-34 on Seen Devices: Per-Class Precision/Recall/F1 and Accuracy (Averages and Ranges Across 60 Devices)

| Metric | Macro Avg. (%) | Weighted Avg. (%) | Min Class (%) | Max Class (%) |
|---|---|---|---|---|
| Precision | 93.8 | 94.5 | 89.2 | 97.6 |
| Recall | 93.5 | 94.6 | 88.7 | 97.9 |
| F1-score | 93.6 | 94.5 | 88.9 | 97.8 |
| Accuracy | **94.6** (overall test accuracy) | | | |

### 4.3. Classifier Backbones and Watermarking

**Implementation details.** To ensure reproducibility and efficiency, we use reproducible seeding with (CUDA Deep Neural Network library) cuDNN, OneCycleLR for smooth convergence, automatic mixed precision (AMP) for throughput, and Focal Loss to balance hard/easy examples. Training/validation splits with early stopping are applied, and per-class precision/recall/F1 and confusion matrices are saved for auditability. All backbones share the same input pipeline (Section III-B) with label smoothing, AdamW, cosine LR decay, and LightSpecAug.

ResNet-34 offers the best trade-off, reaching 94.6% accuracy with stable convergence and strong generalization. Shallow CNNs underfit (87.4%); ResNet-18 improves to 91.8%; a custom ResNet-35 gains slightly (94.9%) at extra cost; and a hybrid distance head attains 92.1% but complicates watermarking.

**Backbones evaluated:**

- Shallow CNN (3–5 conv layers),
- ResNet-18,
- **ResNet-34 (default)**,
- Custom ResNet-35 (slightly wider),
- Hybrid feature extractor + distance head.

### 4.4. Fine-Grained Evaluation of ResNet-34

Table 3 summarizes per-class metrics. Averages over 60 devices show balanced precision/recall, with minimum recall above 88%. Errors concentrate among look-alike emitters, motivating augmentation and anomaly-guard screening.

**Table 4**. Watermark Strategies vs. Attacks: Efficacy and Verification Metrics

| Watermark | Clean Acc. (%) | ASR (%) | Resists Prune/Quantize | Resists Sanitization | Verification Handle |
|---|---|---|---|---|---|
| Trigger | 94.5 | 98.7 | Medium (if layers intact) | Medium (drops w/o adv. training) | Query with trigger→$y_{wm}$ |
| Adversarial trigger | 94.4 | 97.9 | Medium | **High** (survives filtering/denoise) | Query with perturbed trigger |
| Gradient/parameter | 94.3 | N/A | **High** (signature persists) | High (input-agnostic) | Penultimate feature/weight signature check |

### 4.5. Watermark Strategies vs. Attacks

We evaluate three watermark designs—deterministic trigger, adversarially trained trigger, and gradient/parameter signature—under the same training schedule as ResNet-34. Table 4 aggregates both task metrics (clean accuracy) and verification metrics (attack success rate, robustness). Trigger marks provide strong black-box proof (ASR≈99%) with negligible accuracy cost; adversarial triggers remain effective after denoising; gradient/parameter signatures persist through pruning/quantization and fine-tuning.

## 5. CONCLUSION

We presented a framework for radio frequency fingerprint identification (RFFI) that integrates a fixed STFT–Mel front end, a ResNet-34 classifier, layered watermarking, and a model-agnostic autoencoder guard. The goal was to achieve high classification accuracy while defending against model theft, weight tampering, and input-space evasion.

On the LoRa dataset, **ResNet-34** emerged as the most practical backbone, reaching 94.6% accuracy with stable convergence and strong generalization, outperforming shallower CNNs and ResNet-18. We showed that implementing **three watermarking schemes**—trigger, adversarial trigger, and gradient/parameter signatures—provides robust ownership verification without degrading accuracy. Triggers achieved nearly 99% success in black-box settings, adversarial triggers remained effective under noise and filtering, and gradient signatures persisted even after pruning and quantization. Together, these mechanisms address input-level and weight-level attacks. In parallel, the **ConvVAE guard**, trained with KL warm-up and free-bits, flagged off-manifold inputs, sanitized triggers, and adversarial perturbations with AUROC 0.90 and AP 0.88, while preserving ~93% of benign traffic. This highlights the value of anomaly detection as a complementary safeguard to watermarking.

Overall, our results show that coupling ResNet-34 with layered watermarks and a robust ConvVAE yields an accurate, secure, and verifiable RFFI pipeline, suitable for IoT authentication and mission-critical wireless deployments. Future work could also explore extending watermarking methods to emerging hardware platforms such as *memristive crossbar arrays*, as demonstrated in [16], to further strengthen system robustness.

## 6. ACKNOWLEDGEMENTS

The research was sponsored by the Army Research Laboratory and was accomplished under Cooperative Agreement Number W911NF-23-2-0014. (ARL-KRI FREEDOM Project: Analog Edge Processor with Embedded Security for Low-power Low-latency Sensing Applications)